\documentclass[twocolumn]{aastex631}

\usepackage{amsmath}

\begin{document}

\title{Phase mixing of propagating Alfv\'{e}n waves in a single-fluid partially ionized solar plasma}

\author[0009-0001-1997-4361]{M. McMurdo}
\correspondingauthor{M. McMurdo}
\affiliation{Plasma Dynamics Group, School of Mathematics and Statistics, The University of Sheffield, Hicks Building, Hounsfield Road, Sheffield, S3 7RH, UK}
\email{mmcmurdo1@sheffield.ac.uk}
\author[0000-0002-3066-7653]{I. Ballai}
\affiliation{Plasma Dynamics Group, School of Mathematics and Statistics, The University of Sheffield, Hicks Building, Hounsfield Road, Sheffield, S3 7RH, UK}
\author[0000-0002-9546-2368]{G. Verth}
\affiliation{Plasma Dynamics Group, School of Mathematics and Statistics, The University of Sheffield, Hicks Building, Hounsfield Road, Sheffield, S3 7RH, UK}
\author[0000-0003-3659-8000]{A. Alharbi}
\affiliation{Department of Mathematics, Jamoum University College, Umm Al-Qura University, Jamoum, 25375 Makkah, Saudi Arabia}
\author[0000-0002-0893-7346]{V. Fedun}
\affiliation{Plasma Dynamics Group, Department of Automatic Control and Systems Engineering, The University of Sheffield, Mappin Street, Sheffield, S1 3JD, UK}



\begin{abstract}

Phase mixing of Alfv\'{e}n waves is one of the most promising mechanisms for heating of the solar atmosphere. The damping of waves in this case requires small transversal scales, relative to the magnetic field direction. Here this requirement is achieved by considering a transversal inhomogeneity in the equilibrium plasma density profile. Using a single fluid approximation of a partially ionized chromospheric plasma we study the effectiveness of the damping of phase mixed shear Alfv\'en waves and investigate the effect of varying the ionization degree on the dissipation of waves. Our results show that the dissipation length of shear Alfv\'en waves strongly depends on the ionization degree of the plasma, but more importantly, in a partially ionized plasma, the damping length of shear Alfv\'en waves is several orders of magnitude shorter than in the case of a fully ionized plasma, providing evidence that phase mixing could be a large contributor to heating the solar chromosphere. The effectiveness of phase mixing is investigated for various ionization degrees, ranging from very weakly to very strongly ionized plasmas. Our results show that phase mixed propagating Alfv\'en waves in a partially ionized plasma with ionization degrees in the range $\mu = 0.518 - 0.657$, corresponding to heights of $1916 - 2150$ km above the solar surface, can provide sufficient heating to balance chromospheric radiative losses in the quiet Sun.

\end{abstract}



\section{Introduction} \label{sec:intro}

The problem of plasma heating in the solar atmosphere is one of the most enigmatic questions that still eludes a definite answer. The formation conditions of many spectral lines observed by various instruments show that the temperature in the solar atmosphere undergoes a gradual increase from the temperature minimum in the photosphere to the chromosphere, to a very steep increase in the narrow transition region and reaches values of several million degrees in the solar corona. The energy in the upper part of the solar atmosphere is lost due to radiation and this needs to be compensated by mechanisms that can balance the radiative loss. 

Several proposed mechanisms can generate the required energy in the chromosphere and/or corona \citep[see, e.g.][]{Erdelyi2007}, however, none of these received unequivocal observational proof. All the proposed mechanisms are likely to work simultaneously under different circumstances. 

One category of heating mechanisms involves energy derived from damping magnetohydrodynamic (MHD) waves. Such waves are generated by various sources (e.g. buffeting of magnetic structures by granular motion) in the dense solar photosphere, propagate along the magnetic field, and can dissipate their energy. Although waves can lose their energy due to, e.g., viscosity, magnetic resistivity, thermal conduction, etc., these effects are not very efficient, as they require long temporal and spatial scales. It is widely accepted that to have an effective energy transfer from MHD waves to smaller scales, one needs large transversal gradients in the background plasma over small spatial scales, e.g. in the ambient Alfv\'{e}n speed. In such conditions, the mechanisms of resonant absorption and phase mixing became ideal wave-based heating candidates.

Resonant absorption, first proposed by \citet{Ionson1978} as a heating mechanism for coronal loops is based on the fact that in a plasma with transversal inhomogeneity, the spectrum of waves becomes continuous. An effective wave energy transfer occurs between incident waves and the plasma if the frequency of the wave lies in the frequency continuum of the plasma. In this situation, the energy of the wave is transferred to the local perturbations in the inhomogeneous regions of the magnetic structure, leading to a growth in the amplitude \citep[see, e.g.,][etc.]{Davila1987, Sakurai1991a, Sakurai1991b, Goossens1995, GoossensRuderman1995, Ballai1999, CallyAdries2010}. The growth of the amplitude leads to nonlinear behaviour of waves near the resonant point that can be balanced by dissipative processes that do not need to be excessively large for resonant absorption to work. The theory of resonant absorption predicts a damping time of kink waves that is proportional to the period of waves and was proposed by \citet{RudermanRoberts2002} to explain the rapid damping of kink oscillations of coronal loops. For a review of the theory of resonant absorption of kink waves see, e.g., \citet{Goossens2011}.  

Phase mixing is a mechanism, like resonant absorption, that requires transversal inhomogeneity, however, it operates based on a different consideration, \citep[see, e.g.,][etc.]{HeyvaertsPriest1983, IrelandPriest1997, Mocanu2008, ProkHood2019, RudermanPetrukhin2017, RudermanPetrukhin2018}. In particular, for the phase mixing of Alfv\'en waves to occur they require an inhomogeneity of the local Alfv\'en phase speed across the background magnetic field. As a result, the perturbations on different magnetic surfaces become out of phase while propagating. As the phase mixing evolves, even with a small amount of viscosity or resistivity, the dissipation mechanisms become important and could eventually transform the wave energy into heat. 

Phase mixing of Alfv\'en waves was first proposed as a possible heating mechanism of the solar corona by \citet{HeyvaertsPriest1983}, who  predicted a damping time proportional to the period of the wave and a damping length inversely proportional to the cube root of the magnetic Reynolds number. Although phase mixing is a viable mechanism that can damp waves, its efficiency (at least under coronal conditions is questionable). Indeed, the studies by \citet{OfmanAschwanden2002} and \citet{Mocanu2008} showed that under coronal conditions the damping length of shear Alfv\'en waves is of the order of a few solar radii, and for an effective damping one needs dissipative coefficients that are several orders of magnitude larger than formulas predict \citep[see, e.g.,][etc.]{Hood2002,Mocanu2008,Cargill2016,PaganoDeMoortel2017,RudermanPetrukhin2018,ProkopyszynHood2019,Vandamme2020}. 

Here we will study the problem of phase mixing for plasma conditions relevant to solar chromosphere that is markedly different from the corona since it is partially ionized, meaning that electrons, positive ions (mostly protons), and neutral atoms interact via long-range collisions (i.e. Coulomb collisions) or short-range encounters (binary interaction of neutrals with other species). These collisions provide a very effective channel for momentum and energy transfer between species but also ensure a certain level of coupling. Based on standard solar atmospheric models, e.g., the AL c7 model \cite{ALC7}, the ratio between the number density of ions to neutrals varies between $10^{-4}$ in the solar photosphere to $10^{2}$ in the upper part of the chromosphere, i.e. over a distance of about 2.5 Mm the plasma changes from being weakly ionized to strongly ionized.  

The modeling framework of a partially ionized plasma depends very much on the frequency regime in which a certain physical mechanism is described. Here we limit ourselves to the case when the frequency of waves is much smaller than the collisional frequency of particles, therefore we employ a single fluid description. In this framework, partial ionization appears only through the generalized Ohm's law and it describes resistive damping (often called ambipolar diffusion or Cowling resistivity) of currents perpendicular to the ambient magnetic field. 

Ambipolar diffusion occurs within a partially ionized plasma when the neutrals are not fully coupled to the charged component. While the charged fluid is subject to the Lorentz force, the decoupled neutral particles undergo a Brownian motion, while still being affected by the close-range collisions with ions. These collisions result in frictional effects between the two components, providing a mechanism for magnetic and mechanical energy to be dissipated, and hence creating a source of localized atmospheric heating \citep[see, e.g.,][]{Khomenko2018, Forteza2007, Shelyag2012}. Studies by \citet{Khomenko2018} and \citet{PopKeppens2021} demonstrated that ambipolar diffusion greatly enhances the damping of waves in photospheric and chromospheric regions where the magnetic field is strong.

The last decade has seen an increased number of studies on the properties of waves in partially ionized plasma. \citet{Zaqarashvili2011} and \citet{Soler2013} discussed the solutions of the dispersion relation for Alfv\'en waves in a two-fluid plasma and found that high-frequency waves (with frequencies higher than the ion-neutral collisional frequency) have vastly different damping rates than the single-fluid model's approach. These studies showed that the efficiency of damping increases for smaller wavelengths. The nonlinear propagation of waves in a three-dimensional stratified solar flux tube in the presence of ambipolar diffusion was considered by \citet{Shelyag2016} and showed that up to $80$\% of the Poynting flux associated with these waves can be dissipated and converted into heat, providing an order of magnitude larger energy supply to the chromosphere compared to the dissipation of stationary currents modelled by \citet{KhomenkoCollados2012}. \citet{Ballai20192} investigated the formation of dispersive shock waves in partially ionized plasmas, in the presence of ambipolar diffusion. These authors showed that the width of shocks increases with the number of neutrals in the system. 

The phase mixing of Alfv\'en waves in a partially ionized plasma, to the best of our knowledge, has so far been neglected. Hence, the present study aims to address this shortcoming by investigating the efficiency of phase mixed Alfv\'en waves in a partially ionized chromospheric plasma and how the damping length of waves depends on the ionization degree of the plasma. The paper is organized as follows: the physical and mathematical background behind the mechanism of phase mixing is presented in Section \ref{sec:PhysicalConsiderations}, where we derive the governing equations for phase mixed Alfv\'en waves in partially ionized plasmas. These equations are solved analytically for a simplified case and numerically for a general setup and a sinusoidal driver in Sections \ref{sec:WeakSol} and \ref{sec:StrongSol}. Finally, our results are summarised and concluded in Section \ref{sec:Conclusions}.

\section{Physical considerations and the governing Alfv\'en wave equations} \label{sec:PhysicalConsiderations}

Our analytical analysis is analogous to the model of  \citet{HeyvaertsPriest1983}, but for a partially ionized, instead of a fully ionized plasma, and the numerical investigation will be similar to the methodology of \citet{Hood2002}. 

\begin{figure}[htp]
   \centering
    \includegraphics[width=.45\textwidth]{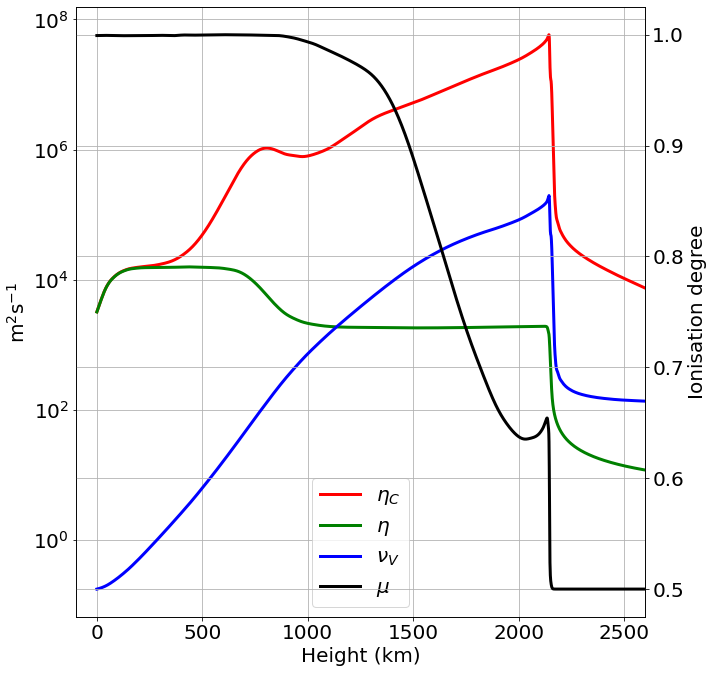}
    \caption{The variation of the ionization degree with height based on the AL c7 atmospheric model \citep{ALC7} (black line) together with the variation of the Spitzer resistivity (green line), Cowling resistivity (red line) and shear viscosity (blue line). These transport coefficients were calculated taking into account the values of the physical parameters given by the AL c7 model.}
  \label{fig:Dissipative_coeffs}  
\end{figure}

Since the temperature of the solar photosphere and chromosphere is not high enough to ensure a full ionization of the plasma (here assumed to be made up of hydrogen) the plasma will consist of a mixture of protons, electrons and neutral atoms that interact through collisions. In order to describe the ionization degree of the plasma, we introduce the relative densities of ions and neutrals (denoted $\rho_{i}$ and $\rho_{n}$, respectively) as 
\begin{equation}
    \xi_{i} = \frac{\rho_{i}}{\rho} \approx \frac{n_{i}}{n_{i} + n_{n}}, \text{   } \xi_{n} = \frac{\rho_{n}}{\rho} \approx \frac{n_{n}}{n_{i}+n_{n}},    \label{eq:Relative densities}
\end{equation}
so that $\xi_i+\xi_n=1$. With these quantities, we can define the ionization degree of the plasma as
\begin{equation}
   \mu = \frac{1}{1 + \xi_{i}},   
    \label{eq:mu}
\end{equation}
which means that $\xi_i=(1/\mu)-1$ and $\xi_n=2-1/\mu$. The quantity $\mu$ will play an important role in our investigation when we discuss the efficiency of the phase mixing in terms of the ionization degree of the plasma. The parameter $\mu$ varies between the values of $1/2$ (fully ionized plasma) to $1$ (fully neutral fluid). In addition, we will use the ratio $\sigma=\xi_i/\xi_n\approx n_i/n_n$ as a measure of the ion number density relative to the neutral number density, so that 
\[
\sigma=\frac{1-\mu}{2\mu-1}.
\]
The degree of ionization of the solar atmospheric plasma can be derived from standard solar atmospheric models. Here we employ the AL c7 model \citet{ALC7} that provides values of neutral hydrogen and electron number densities with height and temperature. The AL c7 model predicts that in the solar photosphere there are $10^4$ more neutrals than ions, while at the top of the chromosphere, ions are more abundant. The ionization degree as a function of height for the AL c7 model is shown in Figure \ref{fig:Dissipative_coeffs}.

In the AL c7 model, the ionization degree is a multi-valued function at heights of about $2.1$ Mm above the solar surface as seen in Figure \ref{fig:Dissipative_coeffs}, i.e. for two different heights we obtain two identical ionization degrees of the plasma. The discrepancies between the values at the top of the chromosphere are due to various non-LTE effects that were considered by the study by \cite{ALC7}. With regard to the present phase mixing study, the different dissipative coefficients will result in different damping lengths for the same ionization degree. However, in our analysis we are going to choose particular values of the ionisation degree that avoid these peculiarities.

The qualitative and quantitative description of the dynamics in a partially ionized plasma depends on the frequency regime of interest. Accordingly, for temporal changes of the same order as the collisional time between particles (or frequencies of the same order as the collisional frequency), an accurate description of the dynamics requires a multi-fluid model. In the present analysis, we assume that the temporal changes are much larger than the collisional time between particles, hence we use a single fluid description, and the partial ionization effects are cast in specific transport mechanisms. 

We consider a plasma permeated by a homogeneous magnetic field oriented in the $z$ direction, and the plasma density varies in the $x$ direction, i.e. $\rho_{0}=\rho_{0}(x)$. We consider a stationary and static equilibrium and the perturbations of the velocity and magnetic field are denoted by $v(x,z,t)\hat{\textbf{y}}$ and $b(x,z,t)\hat{\textbf{y}}$, respectively. Neglecting the effects of gravity and considering the cold and incompressible plasma approximation, the linearised momentum equation is given by

\begin{equation}
    \rho_{0}\frac{\partial \mathbf{v}}{\partial t} = \frac{1}{\mu_0}(\nabla\times {\bf b})\times {\bf B}_0 +\rho_{0}\nu_{v}\nabla^2\mathbf{v},
    \label{eq:NavierS}
\end{equation}
where $\mathbf{B}_0$ is the equilibrium magnetic field oriented in the $z$-direction, $\mu_{0}$ the magnetic permeability of free space and $\nu_{v}$ is the coefficient of viscosity. 
The interaction between the magnetic field and fluid is described by the induction equation that in the linear approximation takes the form
\begin{equation}
    \begin{split}
        \frac{\partial \mathbf{b}}{\partial t} = & \nabla \times (\mathbf{v} \times \mathbf{B}_0) + \eta\nabla^2\mathbf{b} + \\ & \nabla\times\left\{\frac{\eta_{C} - \eta}{|\textbf{B}|^2}\left[\left(\nabla\times\textbf{b}\right)\times\textbf{B}_0\right]\times\textbf{B}_0\right\},  
    \end{split}
  \label{eq:Induction}
\end{equation}
where $\eta$ is the magnetic diffusivity (also known as Spitzer resistivity) and $\eta_{C}$ is the Cowling resistivity. The quantity $\eta_{C}-\eta$ often is replaced by $\eta_{A}$, the coefficient of ambipolar diffusion. Equations (\ref{eq:NavierS}) and (\ref{eq:Induction}) contain a degree of simplifications that will allow us to make analytical progress and the omitted ingredients may play an important role. Gravitational effects are neglected despite that the wavelengths we are going to discuss are large enough that they might cover several gravitational scale-heights. In addition, we will consider that dissipative coefficients are all $z$-independent.

The ambipolar diffusion term in the induction equation is a consequence of considering the effects of neutrals in a single fluid partially ionized plasma and it is due to neutrals not experiencing the magnetic field, and hence, the Lorentz force. As a result, the neutrals become decoupled from the magnetic field and the magnetic field diffuses through the neutral gas without any perturbation caused by the neutrals. For a hydrogen plasma the coefficient of ambipolar diffusion is given by
\begin{equation}
    \eta_{A} = \frac{\xi_{n}^2 v_{A}^2}{\nu_{in}+\nu_{en}},
    \label{eq:Ambipolar}
\end{equation}
where $v_{A}$ is the Alfv\'en speed. The magnetic diffusivity coefficient is dependent on the collisional frequencies of electrons with ions and neutrals and in partially ionized plasma, whose value is given by 
\begin{equation}
    \eta = \frac{m_{e}(\nu_{ei}+\nu_{en})}{e^2n_{e}\mu_{0}}, \label{eq:Spitzer}
\end{equation}
where $m_e$ and $e$ are the electron mass and charge and $n_e$ is the electron number density. In the above relations $\nu_{ab}$ denotes the collisional frequency between species $a$ and $b$ and their expression is given by \citep{Braginskii1965} 

\[
 \nu_{in} = 4n_{n}\sigma_{in}\left(\frac{k_{B}T}{\pi m_{i}}\right)^{\frac{1}{2}}, \quad \nu_{en} = n_{n}\sigma_{en}\left(\frac{8k_{B}T}{\pi m_{e}}\right)^{\frac{1}{2}}, 
 \]
 \begin{equation}
\nu_{ei} = \frac{n_{e}e^4 \Lambda}{3m_{e}^2 \epsilon_{0}^2}\left(\frac{m_{e}}{2\pi k_{B}T}\right)^{3/2}=\nu_{ii}\left(\frac{m_i}{m_e}\right)^{1/2}, 
    \label{eq:CollisionalFreq}
\end{equation}
where $m_i$ is the ion (proton) mass, $\epsilon_0$ is the electric permittivity, $k_B$ is the Boltzmann constant, $T$ is the temperature and 
\begin{equation}
    \Lambda = \ln\left[\frac{8.48\pi}{n_i^{1/2}}\left(\frac{\epsilon_{0}k_{B}T}{e^2}\right)^{3/2}\right]
\end{equation}
is the Coulomb logarithm describing the cumulative effect of many small angle deflective collisions. Here $n_i$ is the ion number density.

The collisional cross-sections that appear in the above relations are height-dependent due to the fact that these values depend on the energy of colliding particles, i.e. their temperature \citep[see, e.g.][]{VranjesKrstic2013}. However, for the sake of the present analysis, we are going to consider them as constant and here these take the values $\sigma_{in}=3.5 \times 10^{-19}$ m$^{2}$, $\sigma_{en} = 10^{-19}$ m$^{2}$.

Since we assume a quasi-neutral hydrogen plasma, $n_e=n_i$. All number densities are given in m$^{-3}$ and temperature in K. As we consider elastic collisions between particles,  the momentum conservation requires that $\nu_{ab}m_an_a=\nu_{ba}m_bn_b$.

Assuming that flow speeds are comparable with the particles’ thermal speed, viscosity can be understood as the flux of momentum. In a plasma, where the direction of the magnetic field defines a preferential direction, the viscosity tensor is more complicated, as the transport of momentum occurs at different rates in different directions. In a magnetically dominated plasma Alfv\'en waves are affected by shear viscosity that originates from random walk transport of momentum with a step size equal to the Larmor radius. 

In a partially ionized plasma, the viscosity of the fluid contains contributions from each species. However, in uni-thermal plasmas, the electron shear viscosity can be negligibly small (proportional to $m_e/m_i$). In addition, under chromospheric conditions the neutral shear viscosity is at least five orders of magnitude larger than the corresponding value for ions, therefore the ion shear viscosity will also be neglected. As a result, the shear viscosity coefficient used in the present study is \citep{vranjes2014}
\begin{equation}
    \nu_{v} = \frac{n_{n}k_BT\tau_{n}}{2}\frac{\Delta\gamma + (\omega_{ci}\tau_i)^2}{\Delta^2+(\omega_{ci}\tau_i)^2},
    \label{eq:shearviscosity}
\end{equation}
where $\omega_{ci}=eB_0/m_i=v_A\left(e^2\mu_0n_i/m_i\right)^{1/2}$ is the ion cyclotron frequency, and
\[
\begin{split}
    & \Delta=1-\frac{1}{(3\nu_{ii}/\nu_{in}+4)(3\nu_{nn}/\sigma\nu_{in}+4)}, \\ & \gamma=1+\frac{\sigma}{3\nu_{ii}/\nu_{in}+4}
\end{split}
\]
with 
\[
\nu_{nn}=4n_n\sigma_{nn}\left(\frac{k_BT}{\pi m_n}\right)^{1/2}
\]
being the neutral-neutral collisional frequency and $\sigma_{nn}$ is the collisional cross-section of this collision, we take this value to be $\sigma_{nn}=2.6\times 10^{-19}$ m$^{2}$.

The variation of the Spitzer resistivity, ambipolar resistivity, and viscosity with height is shown in Figure \ref{fig:Dissipative_coeffs}. Specific plasma parameters were taken from the AL c7 atmospheric model \citep{ALC7}, and the magnetic field varies with height as $B_{0}=1000\exp(-z/660)$ G and $z$ is measured in km. Therefore, the magnetic field used to calculate the dissipative coefficients decays by a factor of $e$ over $660$ km. The governing equation derived by \citet{HeyvaertsPriest1983} was obtained using simplifications such as neglecting terms containing products of dissipative coefficients (in the fully ionised solar corona this assumption might be valid), all terms containing the product of dissipative coefficients and derivatives of equilibrium quantities are neglected and the dissipative coefficients are constant quantities.

In the partially ionized solar chromosphere, the dissipative coefficients are much larger than in the solar corona, therefore the first simplifications made by \citet{HeyvaertsPriest1983} regarding the omission of some terms based on the smallness of dissipative coefficients in the governing equations cannot be applied. Nevertheless, in order to highlight the impact of transport coefficients in partially ionized plasmas and to define a proof-of-concept benchmark, we are going to apply the above simplification. The obtained results could only constitute a qualitative indication of the efficiency of phase mixing in a partially ionized plasma. The proper treatment of the problem will be carried out numerically later. 

\section{Weak solutions} \label{sec:WeakSol}
Let us define the small parameter $\epsilon=l_{inh}/l_{ss}\ll 1$, where $l_{inh}$ is the characteristic spatial scale of inhomogeneity and $l_{ss}$ is a spatial scale of the problem (say, the wavelength of waves). With the inclusion of constant (height-independent) resistivity, viscosity, and ambipolar diffusion, the linearised incompressible MHD equations reduce to
\begin{equation}  
\frac{\partial v}{\partial t} = \frac{B_{0}}{\mu_{0}\rho_{0}}\frac{\partial b}{\partial z} + \nu_{v}\left(\frac{\partial^2 v}{\partial x^2} + \frac{\partial^2 v}{\partial z^2}\right),
\label{eq:LinNav}
\end{equation}
\begin{equation}
  \frac{\partial b}{\partial t} = B_{0}\frac{\partial v}{\partial z} + \eta\left(\frac{\partial^2 b}{\partial x^2} + \frac{\partial^2 b}{\partial z^2}\right) + \eta_{A}\frac{\partial^2 b}{\partial z^2}.
    \label{eq:LinInd}
\end{equation}
We can combine these two equations and eliminate the velocity perturbation, $v$, to obtain a single governing equation describing the evolution of a magnetic field perturbation in the form
\begin{equation}
\begin{split}
\frac{\partial^2 b}{\partial t^2} & =v_{A}^2(x)\frac{\partial^2 b}{\partial z^2} + \\ & +\left[(\eta + \nu_{v})\frac{\partial^2}{\partial x^2} + (\eta_{C} + \nu_{v})\frac{\partial^2}{\partial z^2}\right]\frac{\partial b}{\partial t}+{\cal O}(\epsilon^2),
\label{eq:Gov}
\end{split}
\end{equation}
where the Alfv\'en speed is defined as $v_A(x)=B_0/(\mu_0\rho_{0}(x))^{1/2}$. Similar to the study by \citet{HeyvaertsPriest1983}, we consider that $(\nu_v\partial^2/\partial z^2)/(\nu_v\partial^2/\partial x^2)={\cal O}(\epsilon^2)$, therefore we neglect the viscous term connected to the $z$ derivative. Keeping only the terms of the same order, our governing equation reduces to 
\begin{equation}
\frac{\partial^2 b}{\partial t^2}  = v_{A}^2(x)\frac{\partial^2 b}{\partial z^2} + \left[(\eta + \nu_{v})\frac{\partial^2}{\partial x^2} + \eta_{C}\frac{\partial^2}{\partial z^2}\right]\frac{\partial b}{\partial t}.
\label{eq:Gov_no_viscous_in_x}
\end{equation}

This equation has been obtained using the approximation of  \citet{HeyvaertsPriest1983} that assumes that due to the very high Reynolds numbers as in the solar corona (very small dissipative coefficients), all terms containing products of dissipative coefficients are negligibly small. Due to the dependence of the Alfv\'en speed on the $x$ coordinate, each field line will oscillate with its own frequency since the Alfv\'en speed varies from field line to field line hence we satisfy the conditions for phase mixing to occur.

Following \cite{HeyvaertsPriest1983} we consider that the perturbation of the magnetic field can be written as $b(x,z,t)={\hat b}(x,z)e^{i(\omega t-k_{\parallel}(x)z)}$, where ${\hat b}(x,z)$ is the amplitude of Alfv\'en waves and $k_{\parallel}(x)=2\pi/\lambda_{\parallel}(x)=\omega/v_{A}(x)$ is the parallel wavenumber. After some straightforward calculations, it follows that the amplitude of the magnetic field perturbation can be written as 
\begin{equation}
    \hat{b}(x,z) =  \hat{b}(x,0)\exp\left\{-\left(\frac{z}{\Lambda_{1}}\right)^3-\left(\frac{z}{\Lambda_{2}}\right)\right\},
\label{eq:GovSol2}
\end{equation}
where $\hat{b}(x,0)$ is the amplitude of the perturbation at $z=0$ and the quantities $\Lambda_{1}$ and $\Lambda_{2}$ are related to the waves' damping length. In the above expression the first term in the exponent recovers the results obtained by \cite{HeyvaertsPriest1983}, while the second term is attributed solely to the fact that the plasma is partially ionized. The expressions of these two quantities are given by

\begin{equation}
        \Lambda_{1} = k_{\parallel}^{-1} \left(\frac{6\omega}{(\eta + \nu_{v})\left(d \log k_{\parallel}/dx\right)^2}\right)^\frac{1}{3}, \;\; \Lambda_{2} = \frac{2\omega}{k_{\parallel}^3\eta_{C}}.
\label{eq:LamDamp}
\end{equation}
Hence, Alfv\'en waves will damp due to phase mixing such that the damping length is a superposition of the solution that has the same definition as the expression for the fully ionized case given by \citet{HeyvaertsPriest1983} ($\Lambda_1$), and a term that is due to the ambipolar diffusion ($\Lambda_2$). One very important aspect of these quantities is that only $\Lambda_1$ depends on the gradient of the Alfv\'en speed, while the second one depends on the ionization degree. This qualitatively different behavior is due to the spatial derivatives associated with the dissipative coefficients. While viscosity and Spitzer resistivity are associated with a spatial derivative that is in the direction of propagation, the Cowling resistivity is associated with derivatives perpendicular to the propagation of waves, i.e. perpendicular to the density inhomogeneity. This is a consequence of some of the earlier simplifications. We find that in the numerical case, Cowling resistivity is very much connected to phase mixing.

\begin{figure}[htp]
   \centering
    \includegraphics[width=.45\textwidth]{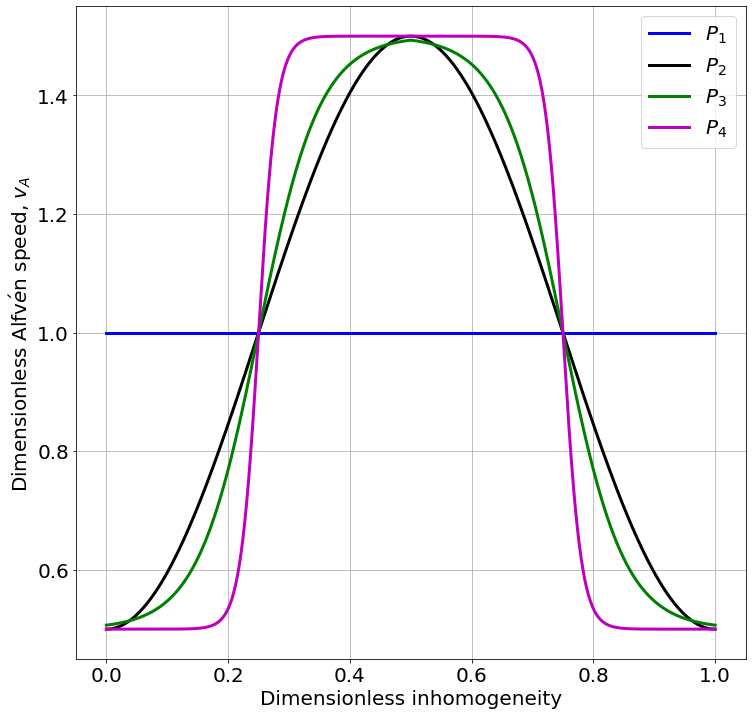}
    \caption{The different profiles of the Alfv\'en speeds used in the analysis are shown by curves of different colors. The constant Alfv\'en speed profile (shown here by the blue horizontal line) will serve as a comparison basis. Speeds and lengths are given in dimensionless units. The two $\tanh$ profiles are symmetric about the midpoint of the inhomogeneity in order to apply periodic boundary conditions to our numerical solver, hence removing the effects of fixed boundaries.}
  \label{fig:vA_profiles}  
\end{figure}

In order to evidence the changes in the damping length of phase-mixed Alfv\'en waves in a partially ionized plasma we perform a simple numerical investigation. For that purpose, we prescribe that the inhomogeneous Alfv\'en speed is given by one the following four profiles
\begin{equation}
    \begin{split}
    & P_1: v_{A}(x) = v_{A1}, \\
    \\ & P_2: v_{A}(x) = v_{A1}\left(1 + \frac{1}{2}\cos\left[\frac{2\pi}{l_{inh}}(x - 0.5l_{inh})\right]\right),
    \\ & P_3: v_{A}(x) = v_{A1}\left(1 + \frac{1}{2}\tanh\left[\frac{x-0.25l_{inh}}{0.1 l_{inh}}\right]\right),
    \\ & P_4: v_{A}(x) = v_{A1}\left(1 + \frac{1}{2}\tanh\left[\frac{x-0.25l_{inh}}{0.03 l_{inh}}\right]\right),
    \end{split}
    \label{eq:vA_profiles}
\end{equation}
where $l_{inh}$ is the length scale of the inhomogeneity. For illustration, we chose $v_{A1}=20$ km s$^{-1}$, $l_{inh} = 300$ km, and the factor of one half is chosen in such a way that the Alfv\'en speed varies by a factor of $3$ across the inhomogeneity, ranging from $10-30$ km s$^{-1}$. The last two $\tanh$ profiles are reflected about the midpoint of the inhomogeneity to make them symmetric. The choice of these profiles was made to model the effectiveness of an increased local gradient in the Alfv\'en speed. These four profiles are converted to dimensionless form before use in the numerical study in Section \ref{sec:StrongSol}, where the reasoning is further discussed relevant to the numerical modeling carried out. Figure \ref{fig:vA_profiles} shows these profiles in dimensionless form.

Since Eq. (\ref{eq:GovSol2}) is not a simple exponential, we cannot define a standard $e$-folding distance. However, we are still going to define the damping length as the length over which the initial amplitude of the wave decays by $e$-times, and this distance will be determined by solving numerically the equation $\left( z/\Lambda_{1}\right)^3 + z/\Lambda_{2}= 1$. The inhomogeneity length scales and characteristic Alfv\'en speeds necessary to perform our investigations were taken from previous studies involving propagating Alfv\'en waves in an inhomogeneous partially ionized plasma in spicules and fibrils which are evidenced in abundance by observations \citep[see e.g.,][etc.]{OberservationHe,ObersvationOkamoto,ObservationBate,ObservationGafeira,ObservationJafarzadeh}. While direct observation of transverse density inhomogeneities are almost impossible to evidence, magnetoconvection codes have shown to produce such transversal density enhancements in fibrils and spicules due to magnetic forces \citep[see e.g.,][etc.]{FibrilFormation,HalphaFormation,SpiculeFormation,SpiculeFormation_AW}.

\begin{figure}[htp]
   \centering 
   \includegraphics[width=.2325\textwidth]{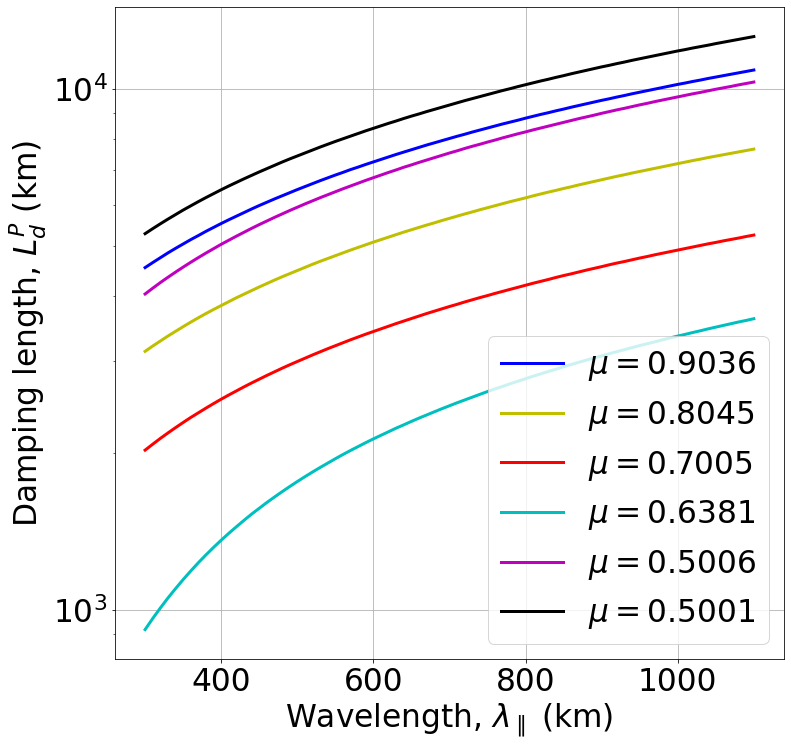}\hfill
    \includegraphics[width=.2325\textwidth]{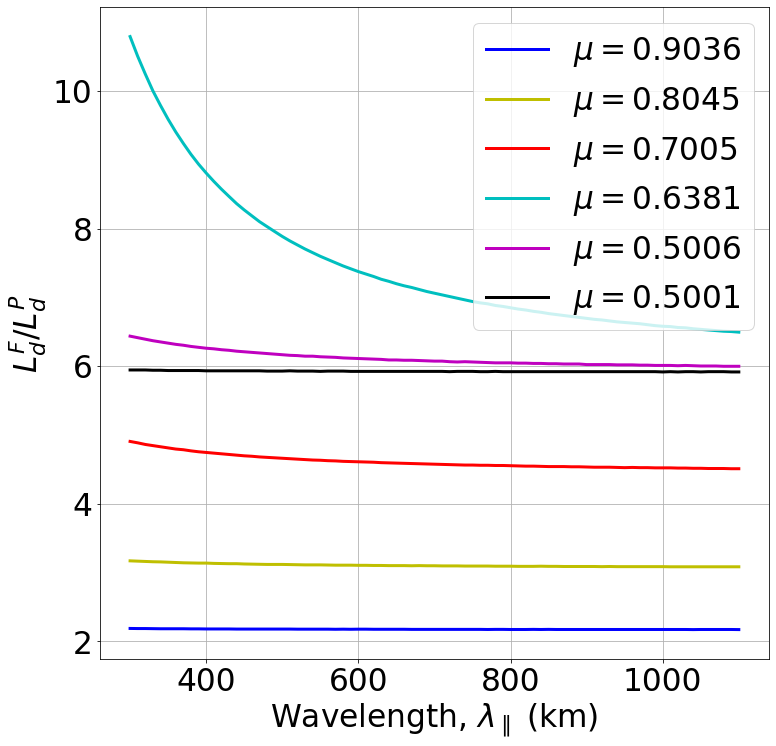}
    \caption{Left: The variation of the damping length of phase-mixed Alfv\'en waves in a partially ionized plasma, $L_{d}^{P}$, in terms of the wavelength of the Alfv\'en waves. Here, and throughout the figures for Section \ref{sec:WeakSol}, unless otherwise stated, the Alfv\'en speed profile is given by the sinusoidal profile in Equation  (\ref{eq:vA_profiles}). Right: The ratio of the damping length of phase mixed Alfv\'en waves in fully ionized, $L_{d}^{F}$, and partially ionized, $L_{d}^{P}$, plasmas. The curves obtained for different ionization degrees are shown by different colors.}
  \label{fig:Weak_analysis}  
\end{figure}
The dependence of the damping length of phase-mixed Alfv\'en waves propagating in a partially ionized plasma on the wavelength of waves is shown in Figure \ref{fig:Weak_analysis} (Left) for different ionization degrees (shown by different colors). The values of the ionization degree were chosen so that these cover the whole spectrum between a strongly ionized and weakly ionized plasma. Our results show that considering either very weakly ionized or very strongly ionized plasma results in weak damping and, therefore large damping lengths, too large to contribute to heating the solar chromosphere. This behavior can be attributed to the variation of the Cowling resistivity with the ionization degree. In both of these cases, the perpendicular currents are very small, thanks to the very small number of neutrals in the case of strongly ionized plasma, and a very high collisional frequency in the case of almost neutral plasma. 

Inspired by these results we have investigated the number of wavelengths an Alfv\'en wave must propagate in order for its amplitude to decay by $e$-times. The results show, again, that the shortest wavelengths are the most affected by phase mixing (due to the relative differences in
speeds of waves in adjacent field lines) and the damping mechanisms considered in our study. For the larger wavelengths, resistivity and viscosity damp the wave most effectively. The largest number of wavelengths is required in the case of nearly fully ionized or neutral plasma. 

The efficiency of phase-mixing in the propagation of Alfv\'en waves in partially ionized plasma becomes visible when we investigate the role of neutrals in this process. For this purpose we consider that by removing entirely the neutral species from the plasma, we have a fully ionized plasma for which the dissipative coefficients will be different and they are given by \cite{HeyvaertsPriest1983}, however the remaining plasma parameters (primarily temperature) remain relevant to the photosphere-chromosphere region. Figure \ref{fig:Weak_analysis} (Right) shows the ratio of the damping lengths obtained in fully ionized and partially ionized plasmas in terms of the wavelength of waves. The different ionization degrees are shown by different colors. Our results show that the ratio in damping lengths is marginally larger for shorter wavelengths meaning the neutrals play a much more important role in wave damping, approximately an order of magnitude reduction in damping when considering the effects of neutrals for ionization degree $\mu = 0.6381$. This behavior is due to the Cowling resistivity found in Eqs. (\ref{eq:GovSol2}) and (\ref{eq:LamDamp}). For short wavelengths (large $k_{\parallel}$), $\Lambda_{2}$ is small and hence is the dominant damping mechanism. For large wavelengths (small $k_{\parallel}$), $\Lambda_{2}$ plays a smaller role and hence for larger wavelengths we find our solutions to tend to a saturated ratio, this is a result of a superposition of solutions coming from the enhanced Spitzer resistivity and shear viscosity coefficients in a partially ionized plasma versus the fully ionized chromospheric plasma, i.e. the damping due to ambipolar diffusion plays a negligible role for large wavelength Alfv\'en waves, at least in the weak solution.

The damping length ratio corresponding to a nearly fully ionized case ($\mu = 0.5001$, shown by the brown line) is independent of the wavelength of waves. In this limit, the contribution due to damping length is due mainly to $\Lambda_1$, and the fact that this ratio is not one can be attributed to the 0.1\% of neutrals still in the system. Furthermore, a similar trend for the weakly ionized cases can be recovered (blue line), that is due to the low values of the Cowling resistivity. These conclusions highlight the need for a balanced population of neutrals and ions in the process of phase mixing.

\begin{figure}[htp]
   \centering
    \includegraphics[width=.45\textwidth]{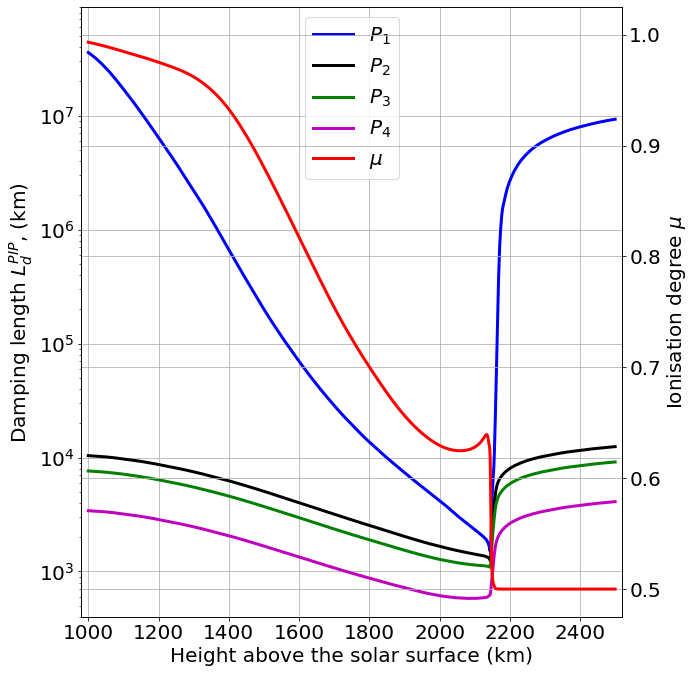}
    \caption{The variation of the damping length with height above the solar surface ranging from $1000-2500$ km in the case of the four Alfv\'en speed profiles ($P_1-P_4$) given by Equation (\ref{eq:vA_profiles}). The particular wavelength used here corresponds to $\lambda_{\parallel} = 400$ km. The overplotted red line shows the ionization degree based on the AL c7 model.}
  \label{fig:Weak_all_vA}  
\end{figure}

Before moving on to the full numerical study, we would like to evidence the effects of varying the Alfv\'en speed profile on the damping lengths of phase-mixed Alfv\'en waves in the ``weak" solution case. We choose to showcase a single wavelength chosen to be $\lambda_{\parallel} = 400$ km and we plot the damping length against all heights associated with ionization degrees in our range of study ($\mu=0.5001 - 0.9036$). The effect of the multi-valued ionization degree is displayed here in Figure \ref{fig:Weak_all_vA}.
There is a clear trend in Figure \ref{fig:Weak_all_vA}, the damping lengths are drastically reduced when the presence of an inhomogeneous Alfv\'en speed is introduced, highlighting the efficiency of phase mixing as a mechanism to damp Alfv\'en waves effectively in the solar chromosphere. The shortest damping length is attained for the steepest profile ($P_4$).

\section{Phase-mixed Alfv\'en waves propagating in a partially ionized plasma: strong solution} \label{sec:StrongSol}

The solution for propagating phase mixed Alfv\'en waves in an unbounded plasma given by Equation (\ref{eq:GovSol2}) can be considered as a ``weak" solution, as it was obtained applying the same simplifications as in the study by \cite{HeyvaertsPriest1983} and they are connected to the small values of dissipative coefficients (or very high Reynolds numbers). While these assumptions are obvious in fully ionized coronal plasmas, in a partially ionized chromospheric plasma, the transport coefficients are large and, therefore, terms containing products of dissipative coefficients have the potential to be not only important but dominant. 

Under these circumstances, the governing equation describing the temporal and spatial evolution of the magnetic field perturbation in a single fluid partially ionized plasma reduces to
\begin{equation}
    \begin{split}
\frac{\partial^2 b}{\partial t^2} = & v_{A}^2(x)\frac{\partial^2 b}{\partial z^2} + \left[(\eta + \nu_{v})\frac{\partial^2}{\partial x^2} + \eta_{C} \frac{\partial^2}{\partial z^2}\right]\frac{\partial b}{\partial t} - \\ & -\nu_{v}\left[\eta\frac{\partial^2}{\partial x^2} + \eta_{C}\frac{\partial^2}{\partial z^2}\right]\nabla^2 b+{\cal O}(\epsilon^2).   \label{eq:Gov_eq_full}
    \end{split}
\end{equation}
In the above equation simplifications were made regarding the order of magnitude difference between characteristic scales in the transversal and longitudinal direction, the second and third terms contain a derivative in the $z$ direction because the Cowling resistivity is much larger in the chromosphere than the Spitzer resistivity and/or viscosity. Given the complexity of the above partial differential equation, the solutions are determined using a numerical method analogous to that used by \cite{Hood2002}. We employ a 4th-order Runge-Kutta-based time step and a 2nd-order centered difference approximation for the spatial derivatives. In order to proceed numerically we must first rewrite Equation (\ref{eq:Gov_eq_full}) as a system of first-order dimensionless equations
\begin{equation}
     \begin{split}
\frac{\partial U}{\partial t} = &  v_{A}^2(x)\frac{\partial^2 b}{\partial z^2} + \left[(\eta + \nu_{v})\frac{\partial^2}{\partial x^2} +\eta_{C} \frac{\partial^2}{\partial z^2}\right]U - \\ & - \nu_v\left[\eta\frac{\partial^2}{\partial x^2} + \eta_{C}\frac{\partial^2}{\partial z^2}\right]\nabla^2 b,
    \end{split}
    \label{eq:GovDimensionless1}
\end{equation}
\begin{equation}
    \frac{\partial b}{\partial t} = U.
    \label{eq:GovDimensionless2}
\end{equation}
Following \cite{Hood2002}, we write all quantities in dimensionless form. Using finite difference approximations for the spatial derivatives to tackle the system of first-order equations \eqref{eq:GovDimensionless1} and \eqref{eq:GovDimensionless2} inevitably leads to solving a large system of linear algebraic equations, which can be written in the form
\begin{align}
    \frac{\partial}{\partial t} \begin{bmatrix}
           b \\
           U
    \end{bmatrix} = A\begin{bmatrix}
           b \\
           U
    \end{bmatrix}
    \label{eq:Tridiagonal}
\end{align}
where $A$ denotes a tri-diagonal matrix with fringes, the coefficients of which correspond to the finite difference approximations. We then use a Runge-Kutta 4th order time step to progress forward in time. The evolution of waves is followed until they reach a steady state or the perturbation reaches the end of the domain. The peaks of the damped wave are tracked and an envelope is fitted. We then calculate the percentage reduction in amplitude after a given distance that the wave has propagated.

\subsection{Sinusoidal wave driver}

When studying the damping of an irregular signal such as a pulse it is informative to study how its constituent components damp. All signals can be approximated by a finite number of differently weighted $sine$ waves represented by the Fourier decomposition of the signal. We can then construct any driver, such as a pulse, from a finite collection of $sine$ waves, each with a definitive damping length. Studying the Fourier decomposition of the bipolar pulse at each time step confirmed that the high-frequency components of the signal damped the fastest as expected.

In this study we consider an infinite sinusoidal wave driver situated at the base of our domain. Understanding what are the main characteristics of waves and how they damp, we can generalise the idea to any driver and quantitatively predict the damping length of waves. That is why extending any of our work to more complicated drivers such as a pulse, broadband or random driver is a simple task.

The  damping of waves is followed for four different Alfv\'en speed profiles specified by Eq. (\ref{eq:vA_profiles}). A constant Alfv\'en speed profile will be chosen as a comparison basis to evidence the direct effects phase mixing has on the damping lengths of the waves. The Alfv\'en speed profiles used in our analysis are shown in Figure \ref{fig:vA_profiles}. 

Introducing an inhomogeneous Alfv\'en speed profile in the transversal direction results in the presence of varying wavelengths in our system. The range of the wavelengths present in our simulation is dictated by the maximum and minimum values of the Alfv\'en speed profile. In our analysis we are going to follow approximately like-for-like wavelengths along the same field line and isolate the effects of a steeper local gradient rather than the effects of a broader range of wavelengths in the simulation. In order to concentrate only on the effect of inhomogeneity we are going to choose the the three inhomogeneous profiles so that their extreme values range between the same dimensionless values $0.5$ and $1.5$.

For simplicity, throughout our analysis, we will follow the modification of the amplitude of Alfv\'en waves on the magnetic field line that corresponds to the intersection of all four profiles displayed in Figure \ref{fig:vA_profiles} occurring at $x = 0.25$. This particular choice removes from the problem the discussion on the effectiveness of damping in terms of the wavelength of waves, however this will be similar to the findings shown in the case of weak solutions. The profiles shown in Figure \ref{fig:vA_profiles} are chosen such that the maximum gradient of the Alfv\'en speed profile occurs at the particular value of $x$. 

\subsection{Results}

Our aim is to study the damping of phase-mixed Alfv\'en waves generated by a sinusoidal wave driver for varying ionization degrees that is given by Eq. (\ref{eq:mu}). As the ionization degree varies between the limits of fully ionized ($\mu=0.5$) and fully neutral ($\mu=1$) plasmas, an increase of the ionization degree is equivalent with the case of a plasma whose temperature is decreasing (assuming that the ionization degree of the plasma is temperature-dependent only, i.e. the effect of radiation is removed). 

\begin{figure}[htp]
   \centering
    \includegraphics[width=.45\textwidth]{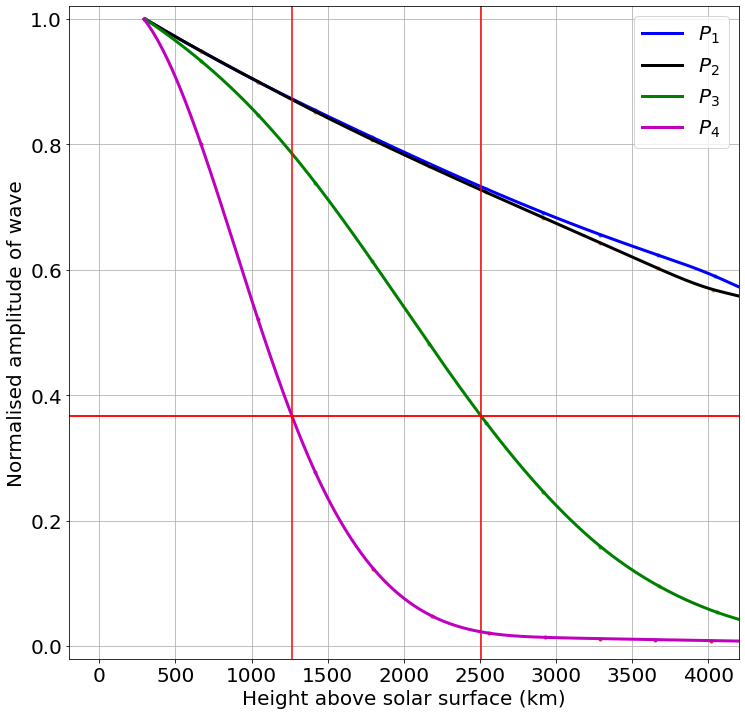}
    \caption{An envelope is fitted to the normalized maxima of Alfv\'en waves in the case of the four Alfv\'en speed profiles. The particular wavelength used in this figure corresponds to  $\lambda_{\parallel}\approx 400$ km and the ionization degree is set to $\mu = 0.7852$.}
\label{fig:Envelope_Wl_400_mu_0.7852}  
\end{figure}

Figure \ref{fig:Envelope_Wl_400_mu_0.7852} shows the normalized value of the envelope in the case of an Alfv\'en wave generated by the four different Alfv\'en wave profiles considered in the present study. The waves have identical periods and the solutions are presented for the same ionization degree ($\mu=0.7852$). The variation in the envelop is caused by different levels of phase mixing due to the variation of the Alfv\'en speed profiles. The envelope is plotted by fitting a curve to the peaks of the generated Alfv\'en waves. The damping length is measured from the first peak, rather than from $z=0$ to avoid any extrapolation errors. The horizontal red line marks the value of the normalized amplitude that corresponds to a $e$-fold decrease of the original value and the vertical lines mark the vertical points where the envelopes of the waves intersect with the horizontal line. The damping length is then calculated to be the distance between the first peak and the vertical line. The results show clearly that any inhomogeneous profile results in a shorter damping length, i.e. phase mixing indeed reduces the damping length of waves. Comparing these results with the profiles of the Alfv\'en waves shown in Figure \ref{fig:vA_profiles}, it is evident that the wave that corresponds to the steepest gradient undergoes the heaviest damping. The steep gradients enhance the contribution of the viscosity and resistivity in the Navier-Stokes and induction equation.

\begin{figure}[htp]
   \centering
    \includegraphics[width=.45\textwidth]{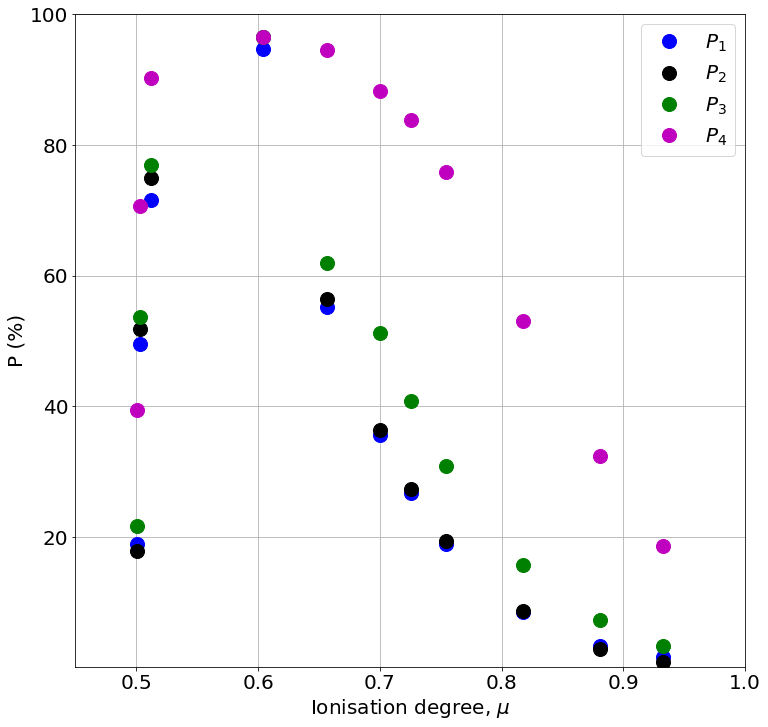}
    \caption{The percentage reduction in the wave amplitude after a propagating length of $1$ Mm is plotted for various ionization degrees, the Alfv\'en speed profiles are labeled in the legend. The steepest of the four profiles gives rise to the most effective wave damping, while all waves are damped effectively for ionization degrees close to $\mu=0.6381$, where the values of viscosity, resistivity, and ambipolar diffusion are at their effective combined maximum. For this figure we study waves with approximate wavelength $\lambda_{\parallel} \approx 400$ km.}
  \label{fig:Percent_damped_dots}  
\end{figure}

The percentage reduction in amplitude of Alfv\'en waves over the particular distance of $1$ Mm from their first peak is displayed in Figure \ref{fig:Percent_damped_dots}. This value was chosen somehow arbitrarily to replace the damping length due to constraints on the size of the domain used in our numerical solver. The percentage reduction of the amplitude is a quantity that is defined as 
\begin{equation}
    \text{P}(\%) = \frac{A[0] - A[1]}{A[0]}\times 100,
    \label{eq:PercentReduction}
\end{equation}
where $A[0]$ is the initial amplitude of the Alfv\'en wave and $A[1]$ is the amplitude after the wave propagated a distance of $1$ Mm. The results shown in Figure \ref{fig:Percent_damped_dots} show the same trend as obtained in the case of weak damping, i.e. a plasma with an ionization degree in the region of $\mu = 0.6381$ produces the most effective damping, where for all the considered Alfv\'en speed profiles nearly all of the wave energy has been dissipated. For larger ionization degrees (higher relative neutral densities) we recover a similar result as shown in Figure \ref{fig:Envelope_Wl_400_mu_0.7852}, i.e. the percentage change in the amplitude of the wave increases with the steepness of the Alfv\'en wave profile.

\begin{figure}
    \includegraphics[width=.2325\textwidth]{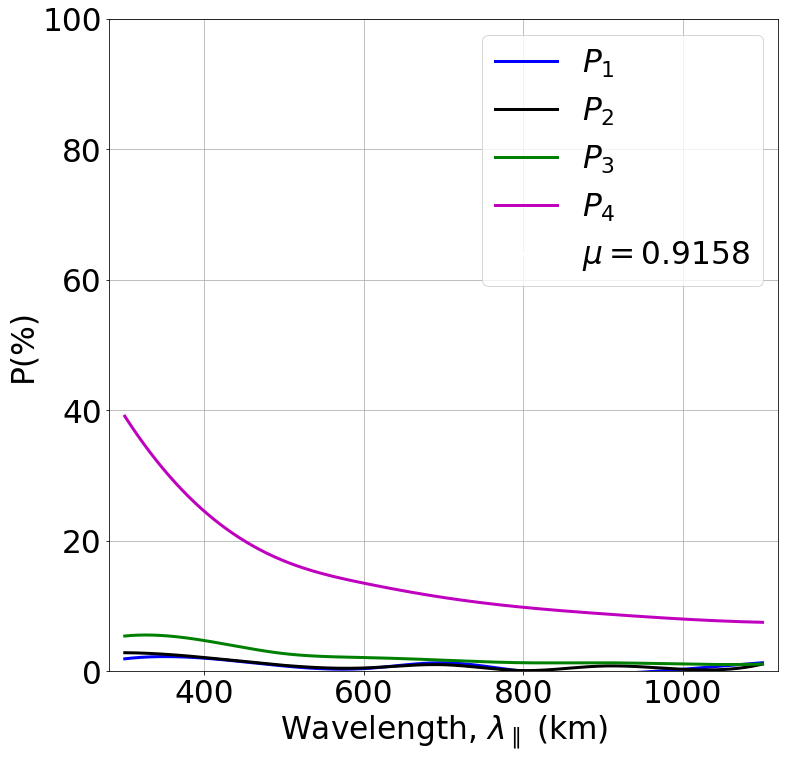}\hfill
    \includegraphics[width=.2325\textwidth]{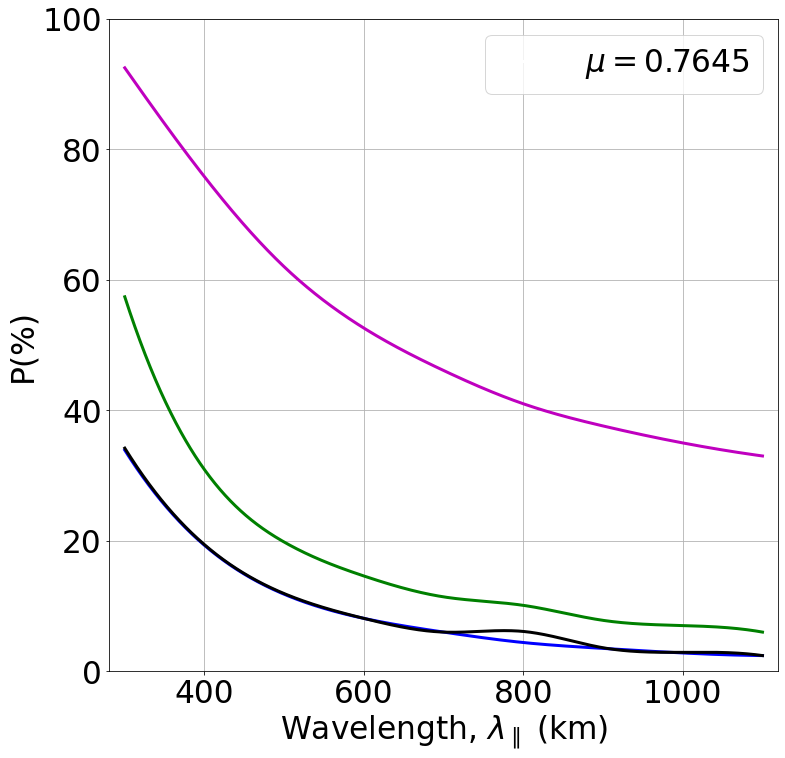}
    \\[\smallskipamount]
    \includegraphics[width=.2325\textwidth]{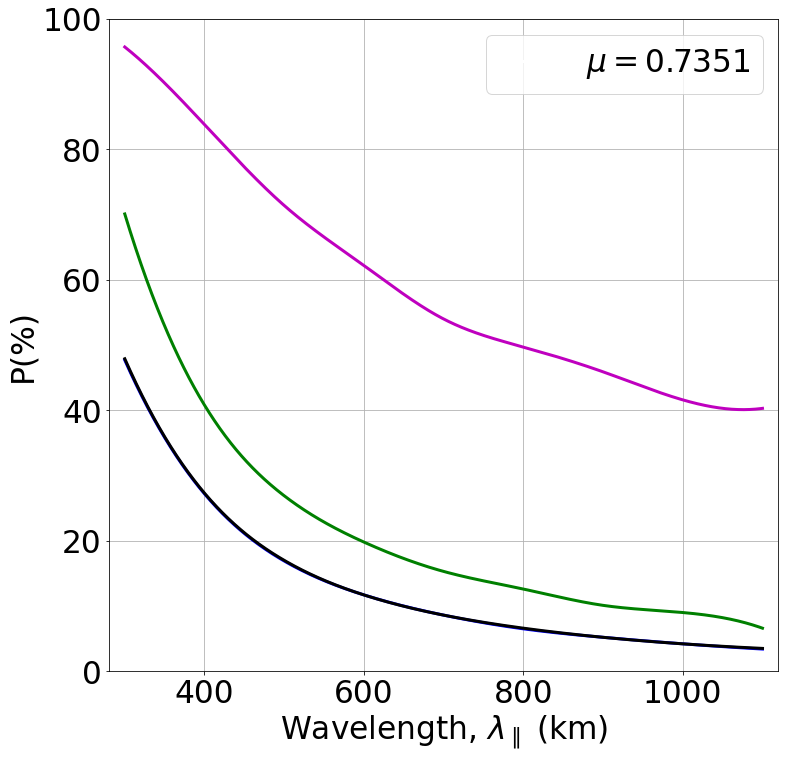}\hfill
    \includegraphics[width=.2325\textwidth]{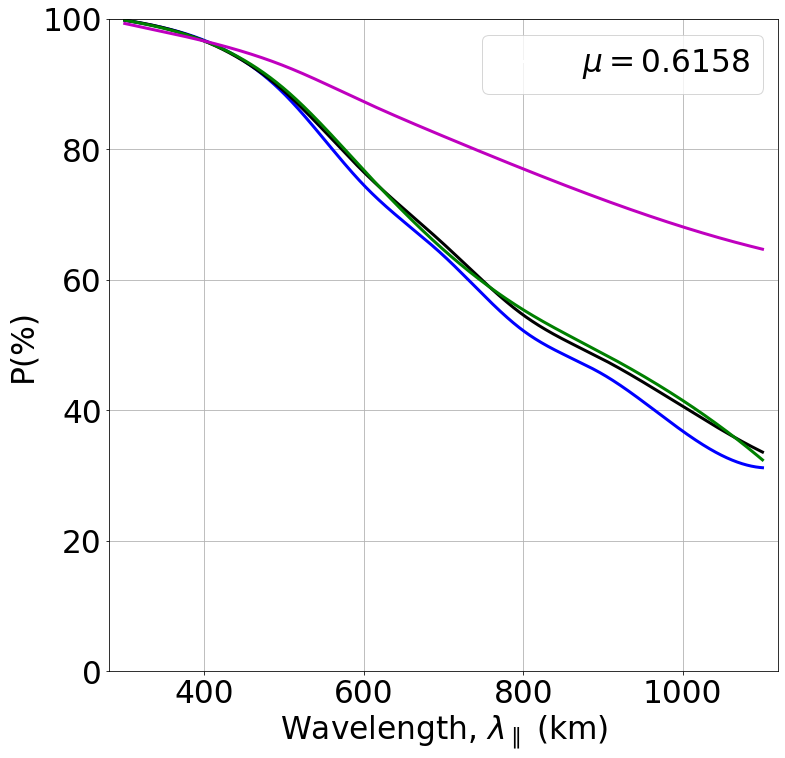}
    \\[\smallskipamount]
    \includegraphics[width=.2325\textwidth]{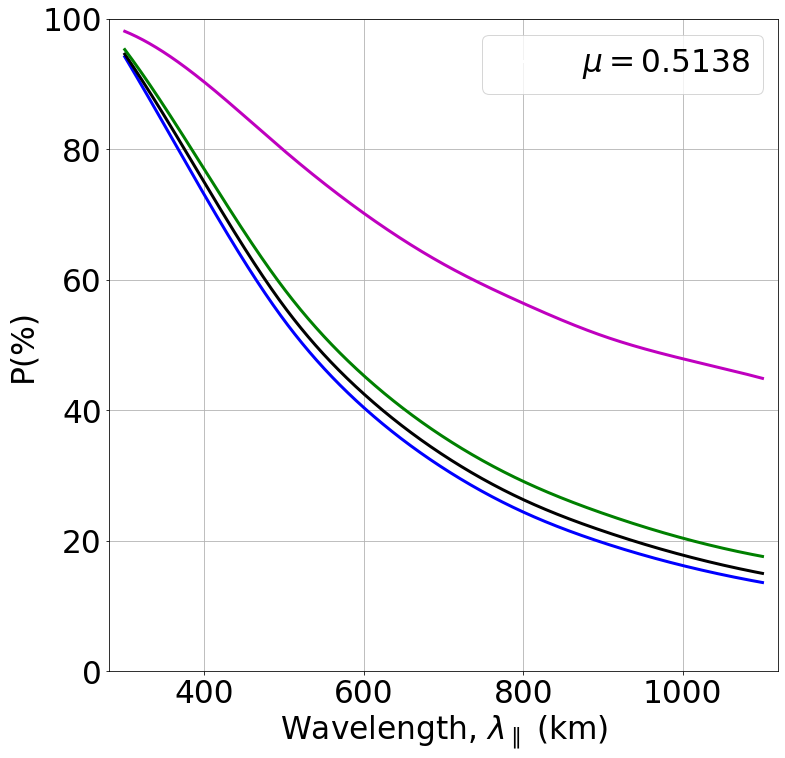}\hfill
    \includegraphics[width=.2325\textwidth]{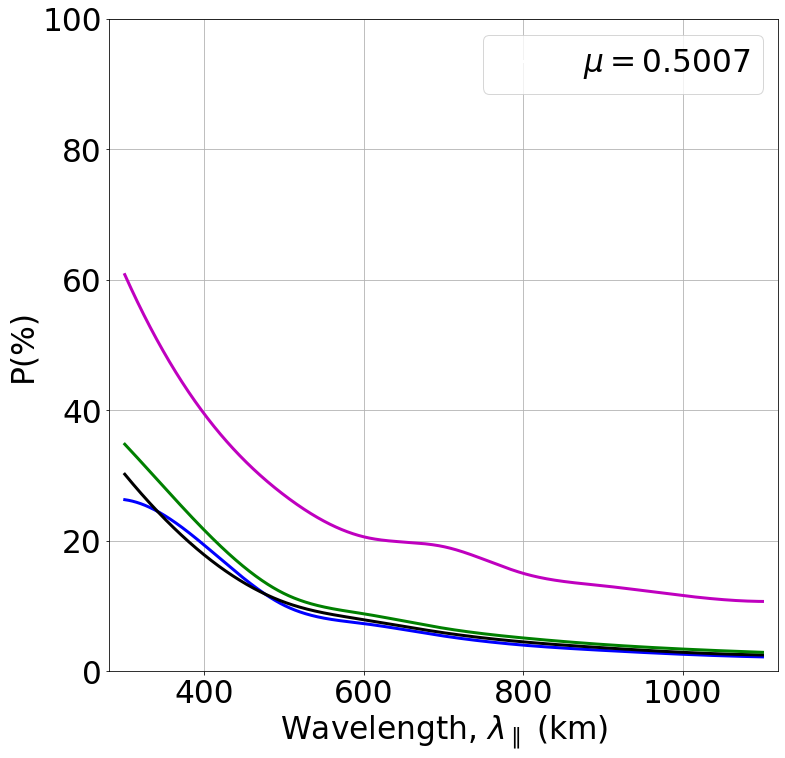}\hfill
    \caption{Percentage change in the amplitude of Alfv\'en waves for six different ionization degrees (displayed in the top right of each panel) in terms of the wavelength of Alfv\'en waves, here in the range of $300-1100$ km. The colors denote the same initial Alfv\'en wave profiles as before.}
    \label{fig:grid_of_6_fullsoln}
\end{figure}
The same percentage damping of Alfv\'en waves after propagating $1$ Mm in terms of the wavelength of waves (here in the range of $300-1100$ km) is shown in Figure \ref{fig:grid_of_6_fullsoln} for different ionization degrees of the plasma. First of all, Alfv\'en waves undergo the smallest damping (irrespective of the wavelength of waves) when the plasma is nearly neutral or nearly fully ionized. Waves propagating in a plasma with an ionization degree close to $\mu = 0.6381$ will damp very effectively for all profiles of Alfv\'en speed as at this value of ionization degree the dissipative coefficients responsible for removing energy from the wave are at their combined largest. Inhomogeneities in the plasma density are shown to enhance the dissipation of Alfv\'en waves in partially ionized plasmas to varying degrees. For each ionization degree considered here, the largest damping belongs to the Alfv\'en wave with the steepest inhomogeneous profile (shown in purple). For ionization degrees close to the fully neutral plasma very little wave energy is dissipated due to resistivity and viscosity when considering the waves that are propagating in a homogeneous plasma, thanks to the weak transport mechanisms. As before, waves with the smallest wavelength damp the most effectively. Our results also show that the damping in the case of a homogeneous plasma (here shown in blue) and in the case of a sinusoidal inhomogeneity lead to almost identical results, meaning that here the damping is not due to the inhomogeneity, instead due to the presence of neutrals.

Let us return to the ``weak" solutions obtained in Section \ref{sec:WeakSol} and verify our hypothesis that the ``strong" solutions are more realistic in chromospheric partially ionized plasmas. In general, taking into account extra terms in the governing equation of phase-mixed Alfv\'en waves leads to a decrease in the damping length by a factor between $1.4-2.5$. The weak solution captures the strong (numerical) solution most effectively for short wavelengths and steep Alfv\'en speed profiles, however, it fails to do so for larger wavelengths. As the wavelength is increased, the weak solution tends towards capturing only $20$\% of the damping after 1 Mm propagated, hence using the weak solution results in an underestimate in the damping by a factor of $5$. These results show that the additional terms containing products of dissipative coefficients in Eq. (\ref{eq:Gov_eq_full}) become important in describing the damping of Alfv\'en waves in inhomogeneous plasmas. That is why, the weak solution has provided us with an order of magnitude estimate for the damping of Alfv\'en waves in inhomogeneous plasmas, however, for accurate description the use of our full numerical solution is imperative.

\begin{figure}[htp]
   \centering
    \includegraphics[width=.45\textwidth]{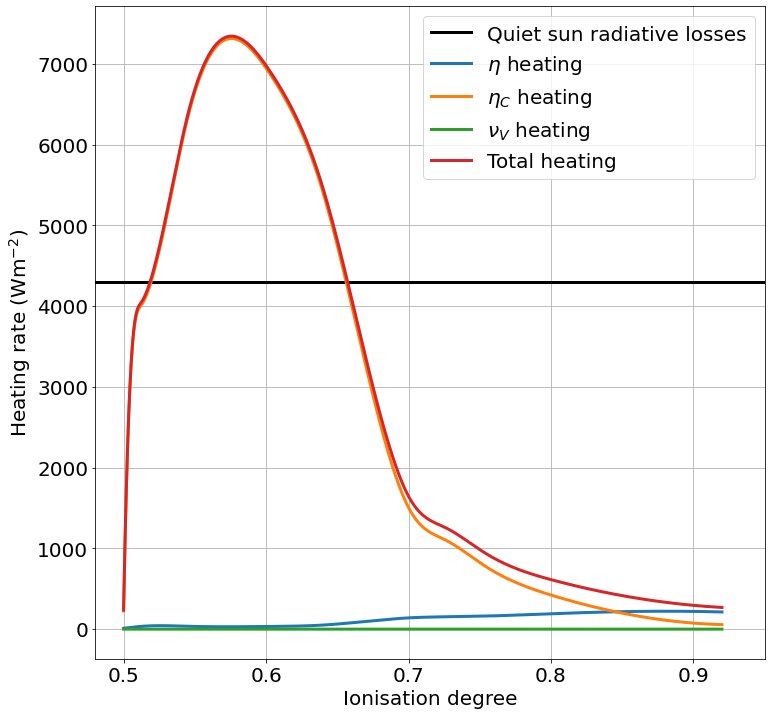}
    \caption{The variation of the heating rate with the ionization degree of the plasma for the phase-mixed Alfv\'en waves with a wavelength of $400$ km described by the $P_4$ profile. Heating rates associated with particular dissipative coefficients are shown by different colours}. The horizontal black line shows the value of the heating rate of the quiet Sun equal to the average radiative losses of the chromosphere.
  \label{fig:Heating_rate}  \end{figure}

Now that we have established that phase mixing in partially ionized solar plasmas has the potential to damp waves very effectively within the chromosphere, let us estimate the amount of heat produced by the damping of phase-mixed Alfv\'en waves that are subject to damping due to Ohmic dissipation of parallel and perpendicular currents, as well as the conversion of waves' kinetic energy into heat due to viscous damping. In this case, the heating rate, $Q$, is calculated as the sum of the heating due to resistive Ohmic heating, $Q_{res}$ and viscous heating, $Q_{\nu}$  \citep{priest,Heating_rate}
\begin{equation}
    \begin{split}
        Q = Q_{res} + Q_{\nu} = & \frac{1}{\mu_{0}}\left[\eta\left(\frac{\partial b}{\partial x}\right)^2 + \eta_{C}\left(\frac{\partial b}{\partial z}\right)^2 \right] + \\ &  + \frac{\rho\nu_{v}}{2}\left(\frac{\partial v}{\partial x}\right)^2,    
    \end{split}
\end{equation}
where we retained only the dominant term in the expression of viscous heating. The variation of the heating rate with respect to the ionization degree of the plasma for a given wavelength of Alfv\'en waves ($400$ km) propagating along the magnetic field line associated with the maximum gradient of the Alfv\'en speed profile $P_4$, shown by purple in Figures \ref{fig:vA_profiles}--\ref{fig:grid_of_6_fullsoln} is displayed in Figure \ref{fig:Heating_rate}. The amplitude of the velocity perturbation was taken to be $2.5$ kms$^{-1}$  \citep{velocity_amplitude}. For the sake of simplicity we neglected the back reaction of the heating process on the values of dissipative coefficients and also the ionisation degree of the plasma. Given the particular dependence of dissipative coefficients, the values we determine constitute an upper limit.

In order to estimate the efficiency of the phase-mixed Alfv\'en waves to heat the plasma, we use, for comparison, the estimated average heating rate of the quiet chromosphere. The radiative losses estimated from commonly used semi-empirical models of the quiet-Sun chromosphere are $4.3$ kWm$^{-2}$, while in active regions, this value reaches $20$ kWm$^{-2}$ \citep{Withbroe1977,Vernazza1981,Radiative_losses}. The required heating rate to compensate for the radiative losses in the quiet chromosphere is shown by the horizontal black line in Figure \ref{fig:Heating_rate}. Our analysis reveals that the maximum heating rate produced by Alfv\'en waves varies considerably with the ionization degree of the plasma, in the full spectrum of the ionization the heating rate varies by more than one order of magnitude and it attains its maximum value for an ionization degree of $\mu=0.576$. The results show that waves propagating in a partially ionized plasma with ionization degrees in the range $\mu=0.518- 0.657$ provide sufficient heating rates to balance chromospheric radiative losses. In the AL c7 atmospheric model, these values correspond to a ratio of neutrals to ions, $n_{n}/n_{i} = 0.0567 - 0.917$, respectively.
\begin{figure}[htp]
   \centering
    \includegraphics[width=.45\textwidth]{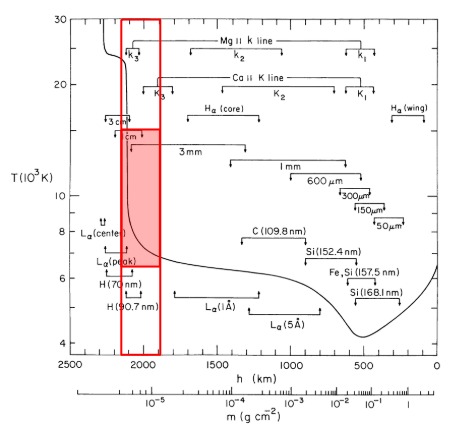}
    \caption{A composite figure showing the formation height and temperature of various continua and spectral lines taken from \citet{Vernazza1981} over which we plot the height and temperature range (shaded box) for which the heating rate produced by phase mixing of Alfv\'en waves is larger than the radiative loses in the quiet chromosphere.}
\label{fig:Spectral_lines}  
\end{figure}
These values of the ionization degree in the AL c7 model occur at the heights of $1916$ and $2150$ km above the solar atmosphere. It is instructive to compare our findings with the variation of the temperature with height in the VAL atmospheric model \citep[][their Figure 1]{Vernazza1981} together with the approximate depths where various continua and spectral lines are formed (see Figure \ref{fig:Spectral_lines}). The heights in between which the heating rate obtained by us is larger than the required heating rate are shown by vertical red lines, while the temperature at which the specific ionization degrees occur is shown by red horizontal segments. In this way, we can define a region in the height-temperature diagram where phase-mixed Alfv\'en waves can provide the required heating. This domain corresponds to the region where the temperature increases dramatically and it is the location where the Lyman$\alpha$ and the 3 mm continuum intensity due mostly to free-free transitions of hydrogen is generated. These results show that the increases in the temperature at the transition region level could be easily attributed to phase mixing of Alfv\'en waves.

Comparing the heating rates obtained using the other three Alfv\'en speed profiles, we found that the ratio in heating rates, obtained by dividing the heating rate associated with $P_4$ by the heating rate from each of the other three profiles, varied considerably with ionization degree. The steepest profile produced as much as $2.5$ times the amount of heating obtained using $P_1$ and approximately $2$ times the heating obtained using the two shallower profiles, $P_2$ and $P_3$. The maximum of this ratio occurs for ionization degrees close to $\mu = 0.6$, the value where we found optimal heating. This is due to the dissipative coefficients reaching their combined maximum close to this ionization degree. For ionization degrees tending towards fully neutral or fully ionized plasmas, this ratio approached unity. This shows the importance of both neutrals and transverse gradients in heating the partially ionized solar chromosphere.

\section{Conclusions} \label{sec:Conclusions}

Ever since the seminal paper by \cite{HeyvaertsPriest1983}, the damping of phase-mixed Alfv\'en waves propagating along magnetic field lines was suggested as a possible mechanism to explain the heating of solar upper atmospheric layers. One key drawback of the phase-mixed Alfv\'en waves applied to the solar corona was that the dissipative coefficients are very small, leading to damping lengths that are often as large as the solar radius. This was attributed to the very small value of transport coefficients in the solar corona.    

Our study aimed to address this major shortcoming by investigating the problem in the chromospheric plasma, where transport coefficients are far larger, however, the plasma is partially ionized. Effective Alfv\'en wave damping due to phase mixing was achieved by imposing a density profile that varied in the transversal direction to the wave propagation. The main result of our research is that using realistic dissipative coefficients for the partially ionized chromospheric plasma results in an enormous reduction in the damping lengths compared to those for the fully ionized coronal case given by \citet{HeyvaertsPriest1983}. Consequently, our estimations do not rely on turbulence to enhance transport mechanisms to bring the damping lengths to values that are important for heating. Small-scale mixing of magnetic field lines will create turbulence that can, even more, enhance transport coefficients \citep{Magyar2017, Turbulence}. 

Our results show that short wavelength Alfv\'en waves damp much faster than longer wavelength waves and the maximum attenuation of Alfv\'en waves occurs for ionization degrees close to 0.6. For steep gradients and for dissipative coefficients corresponding to our optimum ionization degree (or close to this), significant damping is seen for all wavelengths within $1$ Mm of propagation. This length scale of damping could well explain a large amount of chromospheric heating with the rest of the energy left stored in the Alfv\'en waves going on to the corona.

For waves with a particular wavelength of $400$ km propagating in the presence of our steepest profile in the Alfv\'en speed, sufficient heating was generated due to the damping of phase-mixed Alfv\'en waves to balance radiative losses in the upper chromosphere/transition region when the ionization degrees of the plasma is in the range $\mu = 0.518 - 0.657$. This means that phase mixing of Alfv\'en waves in partially ionized plasmas is indeed a viable mechanism for plasma heating.

Finally, we should mention that our approach used several simplifications that made the treatment of the problem of phase-mixing clearer. Key ingredients were neglected (e.g. height and time-dependence of dissipative coefficients, gravitational stratification, etc.), however these might play an important role. Furthermore, as pointed out by \citet{Ofman1998} and \citet{PCargill}, an assumed equilibrium density profile (at least in coronal plasmas) is not sustained due to the heating produced by the damping of phase-mixed Alfv\'en waves. Whether the same conclusion holds for chromospheric partially ionised plasma needs to be investigated by modifying our code to include a time-dependent Alfv\'en speed profile. It is our intention to expand the investigation of this problem and take into account the neglected processes in future analyses.

\section*{Acknowledgements}
 
VF, GV and IB are grateful to the  STFC for grant ST/V000977/1. MM is grateful for the STFC studentship. VF and GV thank the Royal Society, International Exchanges Scheme, collaborations with Pontificia Universidad  Catolica de Chile, Chile (IES/R1/170301) and the Institute for Space-Earth Environmental Research (ISEE, International Joint Research Program, Nagoya University, Japan). VF, GV, IB thank the Royal Society, International Exchanges Scheme, collaborations with Aeronautics Institute of Technology, Brazil, (IES/R1/191114), Monash University, Australia (IES/R3/213012), Instituto de Astrofisica de Canarias, Spain (IES/R2/212183), Institute for Astronomy, Astrophysics, Space Applications and Remote Sensing, National Observatory of Athens, Greece (IES/R1/221095), Indian Institute of Astrophysics, India (IES/R1/211123) and collaboration with Ukraine (IES/R1/211177) for the support provided. This research has also received financial support from the European Union’s Horizon 2020 research and innovation program under grant agreement No. 824135 (SOLARNET). This research was supported by the International Space Science Institute (ISSI) in Bern, through ISSI International Team project 457 (The Role of Partial Ionization in the Formation, Dynamics and Stability of Solar Prominences). The authors are grateful to Referee (Prof P. Cargill) for the constructive comments and suggestions that helped to improve the paper. MM thanks S. McMurdo for the many enlightening conversations that have benefited this research. 
\bibliographystyle{aasjournal}

\bibliography{Paper_1.bib}

\begin{thebibliography}{}
\expandafter\ifx\csname natexlab\endcsname\relax\def\natexlab#1{#1}\fi
\providecommand{\url}[1]{\href{#1}{#1}}
\providecommand{\dodoi}[1]{doi:~\href{http://doi.org/#1}{\nolinkurl{#1}}}
\providecommand{\doeprint}[1]{\href{http://ascl.net/#1}{\nolinkurl{http://ascl.net/#1}}}
\providecommand{\doarXiv}[1]{\href{https://arxiv.org/abs/#1}{\nolinkurl{https://arxiv.org/abs/#1}}}

\bibitem[{{Avrett} \& {Loeser}(2008)}]{ALC7}
{Avrett}, E.~H., \& {Loeser}, R. 2008, \apjs, 175, 229, \dodoi{10.1086/523671}

\bibitem[{{Ballai} {et~al.}(2019){Ballai}, {Forg{\'a}cs-Dajka}, \&
  {Marcu}}]{Ballai20192}
{Ballai}, I., {Forg{\'a}cs-Dajka}, E., \& {Marcu}, A. 2019, Advances in Space
  Research, 63, 1472, \dodoi{10.1016/j.asr.2018.10.024}

\bibitem[{{Bate} {et~al.}(2022){Bate}, {Jess}, {Nakariakov}, {Grant},
  {Jafarzadeh}, {Stangalini}, {Keys}, {Christian}, \&
  {Keenan}}]{ObservationBate}
{Bate}, W., {Jess}, D.~B., {Nakariakov}, V.~M., {et~al.} 2022, \apj, 930, 129,
  \dodoi{10.3847/1538-4357/ac5c53}

\bibitem[{{Braginskii}(1965)}]{Braginskii1965}
{Braginskii}, S.~I. 1965, Reviews of Plasma Physics, 1, 205

\bibitem[{{Cally} \& {Andries}(2010)}]{CallyAdries2010}
{Cally}, P.~S., \& {Andries}, J. 2010, \solphys, 266, 17,
  \dodoi{10.1007/s11207-010-9612-6}

\bibitem[{{Cargill} {et~al.}(2016{\natexlab{a}}){Cargill}, {De Moortel}, \&
  {Kiddie}}]{Cargill2016}
{Cargill}, P.~J., {De Moortel}, I., \& {Kiddie}, G. 2016{\natexlab{a}}, \apj,
  823, 31, \dodoi{10.3847/0004-637X/823/1/31}

\bibitem[{{Cargill} {et~al.}(2016{\natexlab{b}}){Cargill}, {De Moortel}, \&
  {Kiddie}}]{PCargill}
---. 2016{\natexlab{b}}, \apj, 823, 31, \dodoi{10.3847/0004-637X/823/1/31}

\bibitem[{{Davila}(1987)}]{Davila1987}
{Davila}, J.~M. 1987, \apj, 317, 514, \dodoi{10.1086/165295}

\bibitem[{{Erd{\'e}lyi} \& {Ballai}(2007)}]{Erdelyi2007}
{Erd{\'e}lyi}, R., \& {Ballai}, I. 2007, Astronomische Nachrichten, 328, 726,
  \dodoi{10.1002/asna.200710803}

\bibitem[{{Erd{\'e}lyi} {et~al.}(1999){Erd{\'e}lyi}, {Ballai}, \&
  {Ruderman}}]{Ballai1999}
{Erd{\'e}lyi}, R., {Ballai}, I., \& {Ruderman}, M.~S. 1999, in ESA Special
  Publication, Vol.~9, Magnetic Fields and Solar Processes, ed. A.~{Wilson} \&
  {et al.}, 263

\bibitem[{{Forteza} {et~al.}(2007){Forteza}, {Oliver}, {Ballester}, \&
  {Khodachenko}}]{Forteza2007}
{Forteza}, P., {Oliver}, R., {Ballester}, J.~L., \& {Khodachenko}, M.~L. 2007,
  \aap, 461, 731, \dodoi{10.1051/0004-6361:20065900}

\bibitem[{{Gafeira} {et~al.}(2017){Gafeira}, {Lagg}, {Solanki}, {Jafarzadeh},
  {van Noort}, {Barthol}, {Blanco Rodr{\'\i}guez}, {del Toro Iniesta},
  {Gandorfer}, {Gizon}, {Hirzberger}, {Kn{\"o}lker}, {Orozco Su{\'a}rez},
  {Riethm{\"u}ller}, \& {Schmidt}}]{ObservationGafeira}
{Gafeira}, R., {Lagg}, A., {Solanki}, S.~K., {et~al.} 2017, \apjs, 229, 6,
  \dodoi{10.3847/1538-4365/229/1/6}

\bibitem[{{Goossens} {et~al.}(2011){Goossens}, {Erd{\'e}lyi}, \&
  {Ruderman}}]{Goossens2011}
{Goossens}, M., {Erd{\'e}lyi}, R., \& {Ruderman}, M.~S. 2011, \ssr, 158, 289,
  \dodoi{10.1007/s11214-010-9702-7}

\bibitem[{{Goossens} \& {Ruderman}(1995)}]{GoossensRuderman1995}
{Goossens}, M., \& {Ruderman}, M.~S. 1995, Physica Scripta Volume T, 60, 171,
  \dodoi{10.1088/0031-8949/1995/T60/021}

\bibitem[{{Goossens} {et~al.}(1995){Goossens}, {Ruderman}, \&
  {Hollweg}}]{Goossens1995}
{Goossens}, M., {Ruderman}, M.~S., \& {Hollweg}, J.~V. 1995, \solphys, 157, 75,
  \dodoi{10.1007/BF00680610}

\bibitem[{{Grant} {et~al.}(2018){Grant}, {Jess}, {Zaqarashvili}, {Beck},
  {Socas-Navarro}, {Aschwanden}, {Keys}, {Christian}, {Houston}, \&
  {Hewitt}}]{velocity_amplitude}
{Grant}, S. D.~T., {Jess}, D.~B., {Zaqarashvili}, T.~V., {et~al.} 2018, Nature
  Physics, 14, 480, \dodoi{10.1038/s41567-018-0058-3}

\bibitem[{{He} {et~al.}(2009){He}, {Tu}, {Marsch}, {Guo}, {Yao}, \&
  {Tian}}]{OberservationHe}
{He}, J.~S., {Tu}, C.~Y., {Marsch}, E., {et~al.} 2009, \aap, 497, 525,
  \dodoi{10.1051/0004-6361/200810777}

\bibitem[{{Heyvaerts} \& {Priest}(1983)}]{HeyvaertsPriest1983}
{Heyvaerts}, J., \& {Priest}, E.~R. 1983, \aap, 117, 220

\bibitem[{{Hood} {et~al.}(2002){Hood}, {Brooks}, \& {Wright}}]{Hood2002}
{Hood}, A.~W., {Brooks}, S.~J., \& {Wright}, A.~N. 2002, Proceedings of the
  Royal Society of London Series A, 458, 2307, \dodoi{10.1098/rspa.2002.0959}

\bibitem[{{Ionson}(1978)}]{Ionson1978}
{Ionson}, J.~A. 1978, \apj, 226, 650, \dodoi{10.1086/156648}

\bibitem[{{Ireland} \& {Priest}(1997)}]{IrelandPriest1997}
{Ireland}, J., \& {Priest}, E.~R. 1997, \solphys, 173, 31,
  \dodoi{10.1023/A:1004903128146}

\bibitem[{{Jafarzadeh} {et~al.}(2017){Jafarzadeh}, {Solanki}, {Stangalini},
  {Steiner}, {Cameron}, \& {Danilovic}}]{ObservationJafarzadeh}
{Jafarzadeh}, S., {Solanki}, S.~K., {Stangalini}, M., {et~al.} 2017, \apjs,
  229, 10, \dodoi{10.3847/1538-4365/229/1/10}

\bibitem[{{Khomenko} \& {Collados}(2012)}]{KhomenkoCollados2012}
{Khomenko}, E., \& {Collados}, M. 2012, \apj, 747, 87,
  \dodoi{10.1088/0004-637X/747/2/87}

\bibitem[{{Khomenko} {et~al.}(2018){Khomenko}, {Vitas}, {Collados}, \& {de
  Vicente}}]{Khomenko2018}
{Khomenko}, E., {Vitas}, N., {Collados}, M., \& {de Vicente}, A. 2018, \aap,
  618, A87, \dodoi{10.1051/0004-6361/201833048}

\bibitem[{{Leenaarts} {et~al.}(2012){Leenaarts}, {Carlsson}, \& {Rouppe van der
  Voort}}]{HalphaFormation}
{Leenaarts}, J., {Carlsson}, M., \& {Rouppe van der Voort}, L. 2012, \apj, 749,
  136, \dodoi{10.1088/0004-637X/749/2/136}

\bibitem[{{Leenaarts} {et~al.}(2015){Leenaarts}, {Carlsson}, \& {Rouppe van der
  Voort}}]{FibrilFormation}
---. 2015, \apj, 802, 136, \dodoi{10.1088/0004-637X/802/2/136}

\bibitem[{{Magyar} {et~al.}(2017){Magyar}, {Van Doorsselaere}, \&
  {Goossens}}]{Magyar2017}
{Magyar}, N., {Van Doorsselaere}, T., \& {Goossens}, M. 2017, Scientific
  Reports, 7, 14820, \dodoi{10.1038/s41598-017-13660-1}

\bibitem[{{Mart{\'\i}nez-Sykora} {et~al.}(2011){Mart{\'\i}nez-Sykora},
  {Hansteen}, \& {Moreno-Insertis}}]{SpiculeFormation}
{Mart{\'\i}nez-Sykora}, J., {Hansteen}, V., \& {Moreno-Insertis}, F. 2011,
  \apj, 736, 9, \dodoi{10.1088/0004-637X/736/1/9}

\bibitem[{Martínez-Sykora {et~al.}(2017)Martínez-Sykora, Pontieu, Hansteen,
  van~der Voort, Carlsson, \& Pereira}]{SpiculeFormation_AW}
Martínez-Sykora, J., Pontieu, B.~D., Hansteen, V.~H., {et~al.} 2017, Science,
  356, 1269, \dodoi{10.1126/science.aah5412}

\bibitem[{{Melis} {et~al.}(2021){Melis}, {Soler}, \&
  {Ballester}}]{Heating_rate}
{Melis}, L., {Soler}, R., \& {Ballester}, J.~L. 2021, \aap, 650, A45,
  \dodoi{10.1051/0004-6361/202140523}

\bibitem[{{Mocanu} {et~al.}(2008){Mocanu}, {Marcu}, {Ballai}, \&
  {Orza}}]{Mocanu2008}
{Mocanu}, G., {Marcu}, A., {Ballai}, I., \& {Orza}, B. 2008, Astronomische
  Nachrichten, 329, 780, \dodoi{10.1002/asna.200811031}

\bibitem[{{Ofman} \& {Aschwanden}(2002)}]{OfmanAschwanden2002}
{Ofman}, L., \& {Aschwanden}, M.~J. 2002, \apjl, 576, L153,
  \dodoi{10.1086/343886}

\bibitem[{{Ofman} {et~al.}(1998){Ofman}, {Klimchuk}, \& {Davila}}]{Ofman1998}
{Ofman}, L., {Klimchuk}, J.~A., \& {Davila}, J.~M. 1998, \apj, 493, 474,
  \dodoi{10.1086/305109}

\bibitem[{{Okamoto} \& {De Pontieu}(2011)}]{ObersvationOkamoto}
{Okamoto}, T.~J., \& {De Pontieu}, B. 2011, \apjl, 736, L24,
  \dodoi{10.1088/2041-8205/736/2/L24}

\bibitem[{{Oppenheim} {et~al.}(2020){Oppenheim}, {Dimant}, {Longley}, \&
  {Fletcher}}]{Turbulence}
{Oppenheim}, M., {Dimant}, Y., {Longley}, W., \& {Fletcher}, A.~C. 2020, \apjl,
  891, L9, \dodoi{10.3847/2041-8213/ab75bc}

\bibitem[{{Pagano} \& {De Moortel}(2017)}]{PaganoDeMoortel2017}
{Pagano}, P., \& {De Moortel}, I. 2017, \aap, 601, A107,
  \dodoi{10.1051/0004-6361/201630059}

\bibitem[{{Popescu Braileanu} \& {Keppens}(2021)}]{PopKeppens2021}
{Popescu Braileanu}, B., \& {Keppens}, R. 2021, \aap, 653, A131,
  \dodoi{10.1051/0004-6361/202140872}

\bibitem[{Priest(2014)}]{priest}
Priest, E. 2014, Magnetohydrodynamics of the Sun (Cambridge University Press)

\bibitem[{{Prokopyszyn} \& {Hood}(2019{\natexlab{a}})}]{ProkHood2019}
{Prokopyszyn}, A.~P.~K., \& {Hood}, A.~W. 2019{\natexlab{a}}, \aap, 632, A93,
  \dodoi{10.1051/0004-6361/201936658}

\bibitem[{{Prokopyszyn} \& {Hood}(2019{\natexlab{b}})}]{ProkopyszynHood2019}
---. 2019{\natexlab{b}}, \aap, 632, A93, \dodoi{10.1051/0004-6361/201936658}

\bibitem[{{Ruderman} \& {Petrukhin}(2017)}]{RudermanPetrukhin2017}
{Ruderman}, M.~S., \& {Petrukhin}, N.~S. 2017, \aap, 600, A122,
  \dodoi{10.1051/0004-6361/201629892}

\bibitem[{{Ruderman} \& {Petrukhin}(2018)}]{RudermanPetrukhin2018}
---. 2018, \aap, 620, A44, \dodoi{10.1051/0004-6361/201833639}

\bibitem[{{Ruderman} \& {Roberts}(2002)}]{RudermanRoberts2002}
{Ruderman}, M.~S., \& {Roberts}, B. 2002, \apj, 577, 475,
  \dodoi{10.1086/342130}

\bibitem[{{Sakurai} {et~al.}(1991{\natexlab{a}}){Sakurai}, {Goossens}, \&
  {Hollweg}}]{Sakurai1991a}
{Sakurai}, T., {Goossens}, M., \& {Hollweg}, J.~V. 1991{\natexlab{a}},
  \solphys, 133, 227, \dodoi{10.1007/BF00149888}

\bibitem[{{Sakurai} {et~al.}(1991{\natexlab{b}}){Sakurai}, {Goossens}, \&
  {Hollweg}}]{Sakurai1991b}
---. 1991{\natexlab{b}}, \solphys, 133, 247, \dodoi{10.1007/BF00149889}

\bibitem[{{Shelyag} {et~al.}(2016){Shelyag}, {Khomenko}, {de Vicente}, \&
  {Przybylski}}]{Shelyag2016}
{Shelyag}, S., {Khomenko}, E., {de Vicente}, A., \& {Przybylski}, D. 2016,
  \apjl, 819, L11, \dodoi{10.3847/2041-8205/819/1/L11}

\bibitem[{{Shelyag} {et~al.}(2012){Shelyag}, {Mathioudakis}, \&
  {Keenan}}]{Shelyag2012}
{Shelyag}, S., {Mathioudakis}, M., \& {Keenan}, F.~P. 2012, \apjl, 753, L22,
  \dodoi{10.1088/2041-8205/753/1/L22}

\bibitem[{{Soler} {et~al.}(2013){Soler}, {Carbonell}, {Ballester}, \&
  {Terradas}}]{Soler2013}
{Soler}, R., {Carbonell}, M., {Ballester}, J.~L., \& {Terradas}, J. 2013, \apj,
  767, 171, \dodoi{10.1088/0004-637X/767/2/171}

\bibitem[{{Van Damme} {et~al.}(2020){Van Damme}, {De Moortel}, {Pagano}, \&
  {Johnston}}]{Vandamme2020}
{Van Damme}, H.~J., {De Moortel}, I., {Pagano}, P., \& {Johnston}, C.~D. 2020,
  \aap, 635, A174, \dodoi{10.1051/0004-6361/201937266}

\bibitem[{{Vernazza} {et~al.}(1981){Vernazza}, {Avrett}, \&
  {Loeser}}]{Vernazza1981}
{Vernazza}, J.~E., {Avrett}, E.~H., \& {Loeser}, R. 1981, \apjs, 45, 635,
  \dodoi{10.1086/190731}

\bibitem[{{Vranjes}(2014)}]{vranjes2014}
{Vranjes}, J. 2014, \mnras, 445, 1614, \dodoi{10.1093/mnras/stu1887}

\bibitem[{{Vranjes} \& {Krstic}(2013)}]{VranjesKrstic2013}
{Vranjes}, J., \& {Krstic}, P.~S. 2013, \aap, 554, A22,
  \dodoi{10.1051/0004-6361/201220738}

\bibitem[{{Withbroe} \& {Noyes}(1977)}]{Withbroe1977}
{Withbroe}, G.~L., \& {Noyes}, R.~W. 1977, \araa, 15, 363,
  \dodoi{10.1146/annurev.aa.15.090177.002051}

\bibitem[{{Yadav} {et~al.}(2022){Yadav}, {de la Cruz Rodr{\'\i}guez}, {Kerr},
  {D{\'\i}az Baso}, \& {Leenaarts}}]{Radiative_losses}
{Yadav}, R., {de la Cruz Rodr{\'\i}guez}, J., {Kerr}, G.~S., {D{\'\i}az Baso},
  C.~J., \& {Leenaarts}, J. 2022, \aap, 665, A50,
  \dodoi{10.1051/0004-6361/202243440}

\bibitem[{{Zaqarashvili} {et~al.}(2011){Zaqarashvili}, {Khodachenko}, \&
  {Rucker}}]{Zaqarashvili2011}
{Zaqarashvili}, T.~V., {Khodachenko}, M.~L., \& {Rucker}, H.~O. 2011, \aap,
  529, A82, \dodoi{10.1051/0004-6361/201016326}

\end{thebibliography}

\end{document}